\documentclass{aa}
\usepackage{amssymb}
\usepackage{amsmath}
\usepackage{graphicx}
\usepackage{txfonts}
\usepackage{booktabs}
\usepackage{textcomp}
\usepackage{color}
\usepackage{enumitem}
\bibpunct{(}{)}{;}{a}{}{,}
\usepackage{amsfonts,amsmath,graphicx,natbib,color,gensymb,longtable}

\title{A cumulative search for hard X/$\gamma$--Ray emission associated with fast radio bursts in {\em Fermi}/GBM data}
\titlerunning{Cumulative FRB}
\authorrunning{Martone~et~al.}
\author{R.~Martone\textsuperscript{1,2}, C.~Guidorzi\textsuperscript{1}, R.~Margutti\textsuperscript{3}, L.~Nicastro\textsuperscript{4}, L.~Amati\textsuperscript{4}, F.~Frontera\textsuperscript{1,4}, M.~Marongiu\textsuperscript{1,2}, M.~Orlandini\textsuperscript{4}, E.~Virgilli\textsuperscript{1}}
\institute{\textsuperscript{1} Dipartimento di Fisica e Scienze della Terra, Universit\`a di Ferrara, via Saragat 1, I-44122, Ferrara, Italy\\
\textsuperscript{2} ICRANet, Piazza della Repubblica 10, I-65122 Pescara, Italy\\
\textsuperscript{3} Center for Interdisciplinary Exploration and Research in Astrophysics and Department of Physics and Astronomy, Northwestern University, 2145 Sheridan Road, Evanston, IL 60208-3112, USA\\
\textsuperscript{4} INAF--Osservatorio di Astrofisica e Scienza dello Spazio di Bologna, Via Piero Gobetti 93/3, I-40129 Bologna, Italy}

\begin{document}

\abstract 
    {Fast Radio Bursts (FRBs) are millisecond-long bursts uniquely detected at radio frequencies. FRB\,131104 is the only case for which a $\gamma$--ray transient positionally and temporally consistent was claimed. This high-energy transient had a duration of $\sim400$~s and a 15--150~keV fluence $S_{\gamma}\sim4\times10^{-6}$~erg $\mathrm{cm}^{-2}$. However, the association with the FRB is still debated.}
    {We aim at testing the systematic presence of an associated transient high-energy counterpart throughout a sample of the FRB population.}
    {We used an approach like that used in machine learning methodologies to accurately model the highly-variable {\em Fermi}/GBM instrumental background on a time interval comparable to the duration of the proposed $\gamma$--ray counterpart of FRB\,131104. A possible $\gamma$--ray signal is then constrained considering sample average lightcurves.}
    {We constrain the fluence of the possible $\gamma$--ray signal in the 8--1000 keV band down to $6.4 \times 10^{-7}$ ($7.1 \times 10^{-8}$)\,erg\,cm$^{-2}$  for a 200-s (1-s) integration time. Furthermore, we found the radio-to-gamma fluence ratio to be $\eta>10^{8}$\,Jy\,ms\,erg$^{-1}$\,cm$^2$.}
    {Our fluence limits exclude $\sim 94\%$ of {\em Fermi}/GBM detected long gamma--ray bursts and  $\sim 96\%$ of {\em Fermi}/GBM detected short gamma--ray bursts. In addition, our limits on the radio-to-gamma fluence ratio point to a different emission mechanism from that of magnetar giant flares. Finally, we exclude a $\gamma$--ray counterpart as fluent as the one possibly associated with FRB\,131104 to be a common feature of FRBs.}

\maketitle

\section{Introduction}
\label{sec:intro}

Fast Radio Bursts (FRBs) are millisecond-long radio pulses detected around $\sim 1$ GHz. They were serendipitously discovered at the Parkes Radio Telescope \citep{Lorimer07,Thornton13} and are now routinely detected by a number of different facilities (see \citealt{Petroff19_rev} for a review). These observations are collected in the FRB catalogue (\texttt{frbcat}\footnote{\url{http://www.frbcat.org}}; \citealt{Petroff16}), that contains $\sim$ 90 FRBs (as of August 2019): this number is currently dramatically increasing thanks to recent wide-field facilities, with about half sample detected in the latest couple of years.

FRBs come without a direct distance indicator, so this information has to be inferred indirectly from the dispersion measure (DM), that tracks the amount of free electrons along the line of sight. The distance information encoded in the DM contains a degeneracy between the contribution of the intergalactic medium (IGM) and that due to the host galaxy and to the local environment surrounding the FRB source.
Measured DMs are larger than the Galactic contribution (DM$_{\rm MW}$) for all the FRBs of the sample, so the extra-galactic origin is widely accepted. However, the estimates on redshift $z$ have large uncertainties, with the bulk of the population inferred to be at $0.1\la z\la1$. As of August 2019, according to the publicly available information, a direct distance measurement is available only for the repeating burst FRB\,121102 and for the two non-repeating bursts FRB\,180924 and FRB\,190523. 
In the former case, interferometric techniques led to the identification of the host, that turned out to be a dwarf galaxy at $z \simeq 0.193$ \citep{Chatterjee17, Tendulkar17}; in the other two cases the hosts were found to be two luminous, early-type galaxies at respectively $z \simeq 0.32$ and $z \simeq 0.66$ \citep{Bannister19, Ravi19b}.

Additionally, the relatively coarse radio localisation of FRBs of $\sim$ few arcmin severely limits the possibility of multi-wave follow-up, and, as a result, prevents solid inferences on the underlying emission mechanism. Consequently, the number of proposed models dramatically increased in the last years (see \citealt{Platts19} for a review). The ms duration directly implies a small size R $\la 10^{6}$ cm of the radiating region, so FRB sources should be compact objects radiating through a coherent emission process. 
Given their large rotational energies and magnetic fields (up to $10^{15}$~G), and the turbulent environment in which they reside, newborn rapidly rotating neutron stars (NS) or magnetars are considered among the most promising FRB progenitor candidates, even if models involving other compact objects, such as black holes (BHs) and white dwarfs (DWs) are not ruled out (see \citealt{Petroff19_rev, Katz18rev} for reviews). 

The search for signals at non-radio wavelengths ended up with no identified counterpart, with the only exception being the detection ($3.2\sigma$ confidence) of a $400$-s long $\gamma$--ray transient possibly associated to FRB\,131104 reported by \citet[hereafter D16]{DeLaunay16} using data of the Burst Alert Telescope (BAT; \citealt{Barthelmy05}) aboard the {\em Neil Gehrels Swift Observatory} \citep{Gehrels04}. This result raised some scepticism \citep{ShannonRavi17}. Recently, \citet{Cunningham19} performed a systematic search for FRB counterparts in the {\em Fermi}/Gamma-ray Burst Monitor (GBM; \citealt{Meegan09}), {\em Fermi}/Large Area Telescope (LAT; \citealt{Atwood09}), and in the  {\it Swift}/BAT data at the times of the FRBs. Concerning Fermi/{\em GBM} data, the authors focused on the background modelling and subtraction of the individual lightcurves, investigating the presence of a significant excess. They found no evidence for it.

Motivated by these developments, we present here the first cumulative homogeneous search for long-duration $\gamma$--ray emission possibly coincident with FRBs. In particular, we performed a cumulative analysis examining the entire FRB catalogue and exploiting the dozens of FRBs for which contemporaneous Fermi/{\em GBM} data are available.

We organised our paper as follows:
in Sect.~\ref{sec:data_an} we describe the dataset selection and reduction, together with the procedure we used; in Sect.~\ref{sec:res} we outline the results, reporting a special case that required specific analysis; in Sect.~\ref{sec:disc} we discuss our findings compared with D16's results. Finally, in Sect. \ref{sec:conc} we summarise our work and discuss possible future developments.


\section{Data analysis}
\label{sec:data_an}

\subsection{Sample selection}
\label{s:sample}

\begin{table*}
\centering
\caption{List of 38 {\em Fermi}/GBM visible (i.e., not Earth-blocked) FRBs.}
\begin{tabular}{lcrrrrrcccc}
\hline
\hline
FRB & Time (UTC) &  $\alpha_{\rm J2000}$ & $\delta_{\rm J2000}$ & $\theta$ $^{\mathrm{a}}$ & $\phi$ $^{\mathrm{b}}$ & $l_{el}$ $^{\mathrm{c}}$ & $F_{R}$ $^{\mathrm{d}}$ & GBM NaI $^{\mathrm{e}}$ & taken $^{\mathrm{f}}$ & comments $^{\mathrm{g}}$\\
& & $\circ$ & $\circ$ & $\circ$ & $\circ$ & $\circ$ & & units & & \\
\hline
090625 & 21:53:51.4 & 46.95 & -29.93 & 41 & 98 & 45 & 0.07 & 9,0,1 & N & VUO \\
110214 & 07:14:10.4 & 20.32 & -49.79 & 103 & 57 & 54 & 1.74 & 2,5,1 & Y & SXF, HBO \\
110523 & 15:06:19.7 & 326.3 & 0.16 & 103 & 257 & 27 & 0.02 & 8,4,7 & N & VUO \\
110626 & 21:33:17.5 & 315.93 & -44.74 & 79 & 219 & 66 & 0.02 & 8,7,11 & Y & - \\
110703 & 18:59:40.6 & 352.71 & -2.87 & 65 & 263 & 70 & 0.07 & 8,7,3 & Y & - \\
121002 & 13:09:18.4 & 273.7 & -85.2 & 46 & 213 & 22 & 0.08 & 7,6,8 & N & Missing data \\
130628 & 03:58:00.2 & 135.76 & 3.44 & 60 & 356 & 76 & 0.04 & 5,3,1 & Y & SXF, HBO \\
130729 & 09:01:51.2 & 205.34 & -6.0 & 82 & 82 & 14 & 0.12 & 2,10,1 & Y & VUO, HBO \\
131104 & 18:04:11.2 & 101.04 & -51.28 & 117 & 275 & 45 & 0.08 & 4,8,3 & N & VUO \\
150215 & 20:41:41.7 & 274.36 & -4.9 & 61 & 54 & 94 & 0.07 & 1,2,0 & Y & VUO, HBO \\
150418 & 04:29:06.7 & 109.15 & -19.01 & 14 & 232 & 72 & 0.06 & 6,7,0 & Y & - \\
150610 & 05:26:59.4 & 161.11 & -40.09 & 132 & 147 & 3 & 0.05 & 10,11,9 & N & Missing data \\
150807 & 17:53:55.8 & 340.63 & -55.08 & 64 & 207 & 1 & 1.52 & 7,11,8 & Y & - \\
151206 & 06:17:52.8 & 290.35 & -4.13 & 80 & 36 & 83 & 0.03 & 2,5,1 & Y & VUO, HBO \\
160317 & 09:00:36.5 & 118.45 & -29.61 & 59 & 250 & 66 & 0.10 & 7,8,6 & Y & - \\
160608 & 03:53:01.1 & 114.17 & -40.8 & 39 & 268 & 29 & - & 6,7,3 & N & VUO \\
170416 & 23:11:12.8 & 333.25 & -55.03 & 21 & 43 & 83 & 3.26 & 0,1,6 & Y & - \\
170428 & 18:02:34.7 & 326.75 & -41.85 & 71 & 103 & 15 & 1.14 & 10,9,2 & Y & - \\
170707 & 06:17:34.4 & 44.75 & -57.27 & 102 & 351 & 59 & 1.74 & 5,4,3 & Y & VUO, HBO \\
170712 & 13:22:17.4 & 339.0 & -60.95 & 140 & 344 & 20 & 1.78 & 5,4,2 & N & Missing data \\
170827 & 16:20:18.0 & 12.33 & -65.55 & 54 & 85 & 10 & 0.08 & 1,9,0 & Y & - \\
171020 & 10:27:58.6 & 333.75 & -19.67 & 106 & 120 & 47 & 12.60 & 10,9,2 & Y & VUO, HBO \\
171116 & 14:59:33.3 & 52.75 & -17.23 & 56 & 171 & 53 & 2.11 & 9,11,7 & N & VUO \\
171213 & 14:22:40.5 & 54.75 & -10.93 & 88 & 157 & 20 & 4.47 & 11,10,9 & Y & VUO, HBO \\
180110 & 07:34:35.0 & 328.25 & -35.45 & 83 & 34 & 49 & 13.80 & 2,5,1 & Y & VUO, HBO \\
180128.2 & 04:53:26.8 & 335.5 & -60.25 & 60 & 311 & 29 & 2.22 & 3,4,5 & N & Missing data \\
180212 & 23:45:04.4 & 215.25 & -3.58 & 70 & 106 & 27 & 3.22 & 10,9,2 & N & Missing Data \\
180324 & 09:31:46.7 & 94.0 & -34.78 & 53 & 91 & 85 & 2.38 & 9,1,0 & N & Missing Data \\
180525 & 15:19:06.5 & 220.0 & -2.2 & 67 & 161 & 84 & 10.10 & 9,11,10 & N & VUO \\
180725.J0613+67 & 17:59:32.8 & 93.25 & 67.07 & 59 & 24 & 4 & 0.48 & 1,5,0 & Y & Focused analysis \\
180727.J1311+26 & 00:52:04.5 & 197.75 & 26.43 & 72 & 290 & 54 & 0.56 & 4,3,8 & Y & OS by Sco X-1 \\
180729.J1316+55 & 00:48:19.2 & 199.0 & 55.53 & 81 & 297 & 79 & 1.36 & 4,3,8 & Y & - \\
180729.J0558+56 & 17:28:18.3 & 89.5 & 56.5 & 130 & 43 & 22 & 0.36 & 2,5,10 & Y & - \\
180730.J0353+87 & 03:37:25.9 & 58.25 & 87.2 & 117 & 305 & 46 & 2.00 & 4,5,8 & Y & - \\
180801.J2130+72 & 08:47:14.8 & 322.5 & 72.72 & 150 & 147 & 11 & 1.12 & 10,11,2 & N & Missing data \\
180812.J0112+80 & 11:45:32.9 & 18.0 & 80.78 & 13 & 131 & 50 & 0.72 & 0,6,9 & Y & - \\
180814.J1554+74 & 14:20:14.4 & 238.5 & 74.02 & 41 & 62 & 24 & 1.00 & 1,0,9 & Y & - \\
180817.J1533+42 & 01:49:20.2 & 233.25 & 42.2 & 115 & 285 & 48 & 1.04 & 4,8,3 & N & VUO \\
\hline
\end{tabular}

\begin{list}{}{}
\item[$^{\mathrm{a}}$]{azimuth with respect to the payload.}
\item[$^{\mathrm{b}}$]{zenith with respect to the payload.}
\item[$^{\mathrm{c}}$]{elevation upon the Earth limb.}
\item[$^{\mathrm{d}}$]{radio fluence in units of $10^{-16}$~erg~cm$^{-2}$, obtained multiplying the radio fluence density by the band width reported in the \texttt{frbcat}.}
\item[$^{\mathrm{e}}$]{list of the best exposed detectors.}
\item[$^{\mathrm{f}}$]{Whether or not a given FRB was used to build the cumulative lightcurve.}
\item[$^{\mathrm{g}}$]{HBO: Hard Band Only; SXF: Solar X--ray Flare; OS: Occultation Step; VUO: Variability of Unknown Origin; Missing data: {\em Fermi}/GBM data unavailable.
}
\end{list}
\label{tab:general}
\end{table*}

We used {\em Fermi}/GBM-NaI data exploiting its extended band-pass (8-1000 keV) and large field-of-view (2/3 of the whole sky, thanks to its open-sky design), which are both crucial when a cumulative analysis of a (potentially) broad-band signal from poorly localised sources is performed.
We compiled our initial list from the FRB catalogue (as of February 4, 2019) discarding all the events that were Earth-blocked within the {\em Fermi}/GBM field-of-view. In addition, we decided to preliminarily exclude the only two repeating sources, FRB121102 and FRB180814.J0422+73, which in principle could belong to a different class and for which systematic researches of {\em Fermi}/GBM counterparts ended with no candidates (e.g., \citealt{Palaniswamy18, Scholz16, Scholz17}). Our first sample consists of 38 FRBs.
We list in Table~\ref{tab:general}, among other quantities, time, position, and radio fluence of the FRBs of this initial sample.

\subsection{Data reduction}
\label{s:datared}
We calculated the position of each FRB with respect to the GBM payload orientation to identify the best exposed NaI detectors at the FRB time. 
For every FRB, we considered the lightcurves of the three most illuminated detectors and summed them. In particular, we used the {\em Fermi} \texttt{ctime} files, with a resolution of $0.256$~s in 8 energy bins. \texttt{ctime} files were preferred over \texttt{tte} files because we needed both to explore different time intervals around the FRB times  (see below), and to explore times at which no events triggered the GBM (and so no \texttt{tte} files were produced). We retrieved the \texttt{ctime} and \texttt{position history} daily files from the GBM Continuous Data FTP archive \footnote{https://fermi.gsfc.nasa.gov/ssc/data/access/}.

\subsection{Procedure}
\label{s:methodology}
Our search was aimed at identifying long-lived, high-energy signals temporally and spatially coincident with radio detection and common over the FRB population. This was performed searching for significant signals over the background level in the cumulative GBM lightcurves. This means that, unlike other complementary works (e.g. \citealt{Cunningham19}), we do not focus on individual FRBs, but rather on the population as a whole, making no assumptions on the spectral shape.
The search for weak, long-lasting signals in the lightcurves requires accurate knowledge of the background. In the case of GBM, the orbit inclination of $\sim26^\circ$ with respect to the Earth equator makes the background significantly more variable than for an equatorial orbit.
In addition, in the softest energy channels it is necessary to take into account the occultation steps caused by known bright X-ray sources, like the Crab Nebula and Scorpius~X-1.
We addressed this issue through an approach like that used in machine learning methodologies, aimed at modelling and characterising the background along the orbit. 
In particular, the approach was designed to define the following aspects:
\begin{enumerate}[label=\textbf{A.\arabic*}]
    \item \label{itm:first} the size of the time window around the FRB time where we search for the signal;
    \item \label{itm:second} the size of the two time windows (before and the after the FRB time) to be used to fit the background;
    \item \label{itm:third} the algorithm to produce the final background-subtracted, averaged lightcurves.
\end{enumerate}
The implementation in a software tool (e.g. using a neural network model) is foreseen but its description goes beyond the goals of this paper. The following steps summarise the adopted procedure:
\begin{enumerate}
    \item we simulated a sample of non Earth-occulted randomly selected sky positions at random times to act as a training set to define a background-fitting procedure;
    \item we built a second different simulated sample to act as a validation set to test the reliability of our procedure;
    \item we finally applied the background modelling procedure to the FRB dataset.
\end{enumerate}

The sizes of the training and validation sets were the same as that of the FRB one. We used the same data reduction procedure described in Sect.~\ref{s:datared}, focusing on the sum of the lightcurves of the three best exposed detectors for each sample element. Hereafter, we set the origin of the time axes at the FRB epoch. 
The strategy consisted in selecting a time window around $t=0$ (central interval) and then using two intervals (before and after the origin) to fit and interpolate the background (background intervals). We then fitted the background using polynomials, progressively increasing their degree (up to 5) until the p--values for both the $\chi^2$ and the run tests were within the $0.05$--$0.95$ range.
We set the sizes of the background and of the central interval following the analysis of the training sample. The final choice was the result of a trade-off between the need of exploring a long interval around $t=0$ and that of accurately interpolating the background within the central interval.
To define \ref{itm:first} and \ref{itm:second}, we considered two different schemes:
\begin{itemize}
    \item a total time window of 800 s extending from $-400$ to 400~s, with a central interval [$-50$, 150~s] and the background interval spanning [$-400$, $-50$~s] and [150, 400~s]; hereafter, this will be referred to as ``short window'' (SW);
    \item a total time window of 1100~s extending from $-400$ to 700~s, with a central interval [$-50$, 350~s] and the background interval spanning [$-400$, $-50$~s] and [350, 700~s]; hereafter, this will be referred to as ``long window'' (LW).
\end{itemize}

\begin{figure}
\centering
\includegraphics[width=\linewidth]{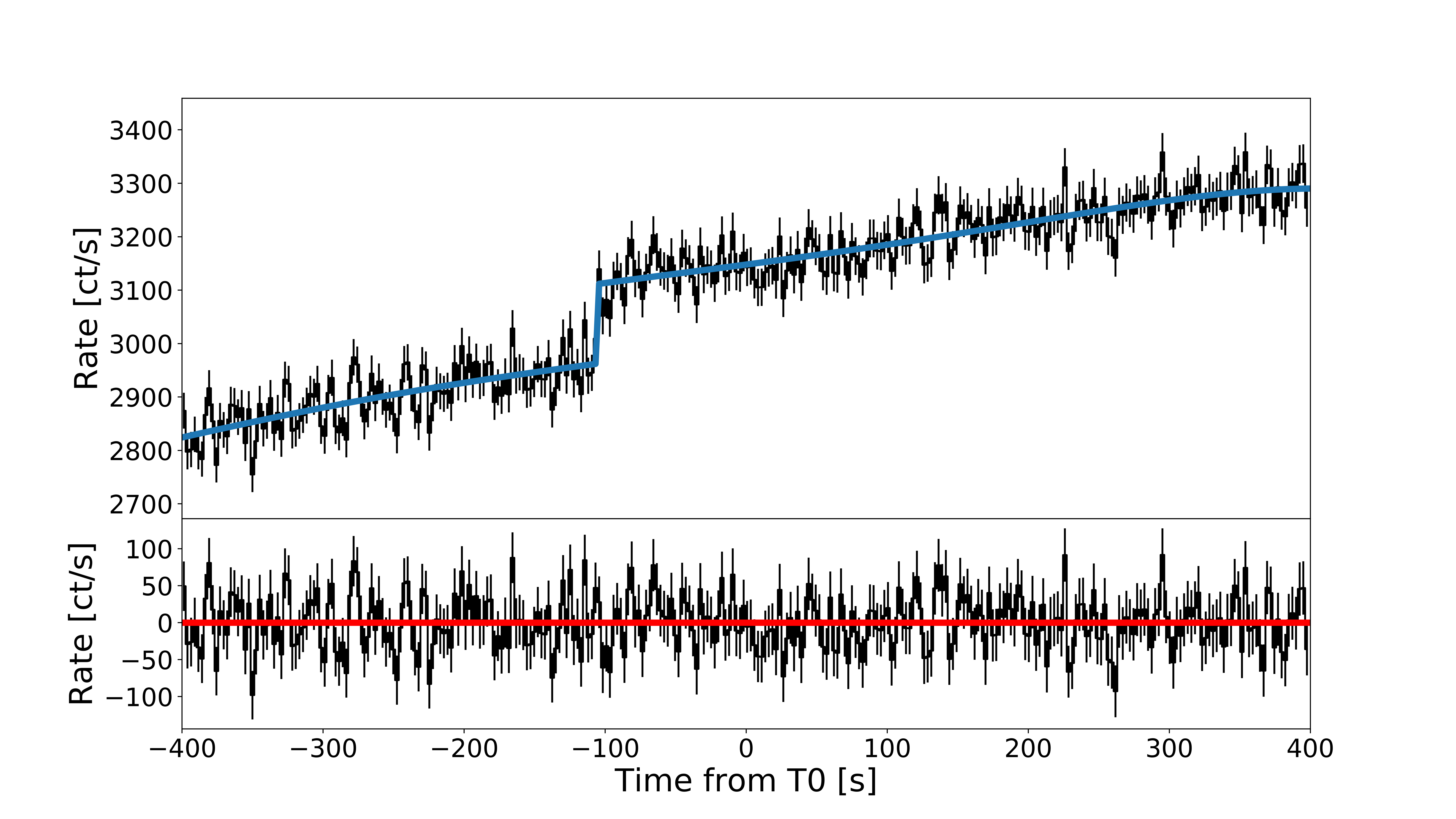}
\caption{Example of step function fitting applied to a case of Sco X--1 occulation step that occurred $\sim 100$~s before FRB180727.J1311+26. {\em Top panel}: 8-1000 keV lightcurve along with the modelled background (solid line). {\em Bottom panel}: residuals of the background modelling.}
\label{fig:step_example}
\end{figure}

Events which passed the tests on the background intervals with both schemes were tagged as good events. 
In case of bad fit possibly caused by occultation steps of bright X/$\gamma$--ray sources, we added a step component to our polynomial fitting function. The result was then tested using the same approach and thresholds already described in the case of a simple polynomial fit.
Figure \ref{fig:step_example} shows the case of FRB180727.J1311+26 which is affected by an occultation step of Sco~X-1.
Several events had to be discarded due to data gaps. In some cases the fit quality did not improve due to unidentified source(s) of variability, which was expected given the number of variable Galactic X--ray sources that contribute to the diffuse background around the lower boundary of the GBM passband. 
Driven by the need to reduce the amount of background contamination sacrificing the smallest possible amount of data, we carried out the same analysis in parallel on a restricted band, ignoring the first two channels of the \texttt{ctime} file, corresponding to selecting $E\gtrsim27$~keV. Hereafter, this will be referred to as the ``hard band''. As expected, the number of poor fits dropped substantially. Overall, given the two background fitting schemes (LW and SW) and the two energy ranges (8-1000 keV and 27-1000 keV), we extracted a maximum of four lightcurves for each element of the training and validation sets. The corresponding individual lightcurves (four lightcurves per FRB) were then averaged over the sample to obtain four average time series.

Concerning \ref{itm:third}, we considered two different schemes:
\begin{itemize}
    \item we averaged the individual background-subtracted lightcurves (hereafter, subtract-and-average strategy);
    \item we performed the background fitting and subtraction on the averaged (non-background subtracted) lightcurves (hereafter, average-and-subtract strategy).
\end{itemize}

We finally evaluated the robustness of the background interpolation procedure by calculating, for each average lightcurve, the significance of the total background-subtracted counts in the central interval. This was done through a polynomial fit of the average lightcurves in the background interval using the same strategy adopted for the individual cases ($\chi^2$ + run tests and $0.05$--$0.95$ thresholds) and finally interpolating in the central interval. Our main results are summarised as follows:
\begin{itemize}
    \item LWs are too long to ensure a reliable interpolation of the background, since we obtained 5-$\sigma$ excesses in the central intervals. This holds true for both the total and the hard band.
    \item SWs allow for a good interpolation of the background, since we obtained no significant excesses ($>3\sigma$) on the central interval in either the training or validation samples for both energy ranges.
    \item The subtract-and-average strategy leads to a 5-$\sigma$ excesses in the central interval, so it does not ensure a reliable interpolation of the background.
    \item The average-and-subtract strategy leads to no significant excesses ($>3\sigma$) on the central interval, so it does ensure a reliable interpolation of the background.
\end{itemize}

Therefore, we hereafter consider only the SW procedure for the analysis of the FRB sample, using the average-and-subtract strategy.
\subsection{An intriguing case: FRB\,180725.J0613+67}
The case of FRB180725.J0613+67 required a dedicated analysis, since it went off during the rise of an occultation step of the Crab nebula.

\begin{figure*}
\centering
\includegraphics[width=\linewidth]{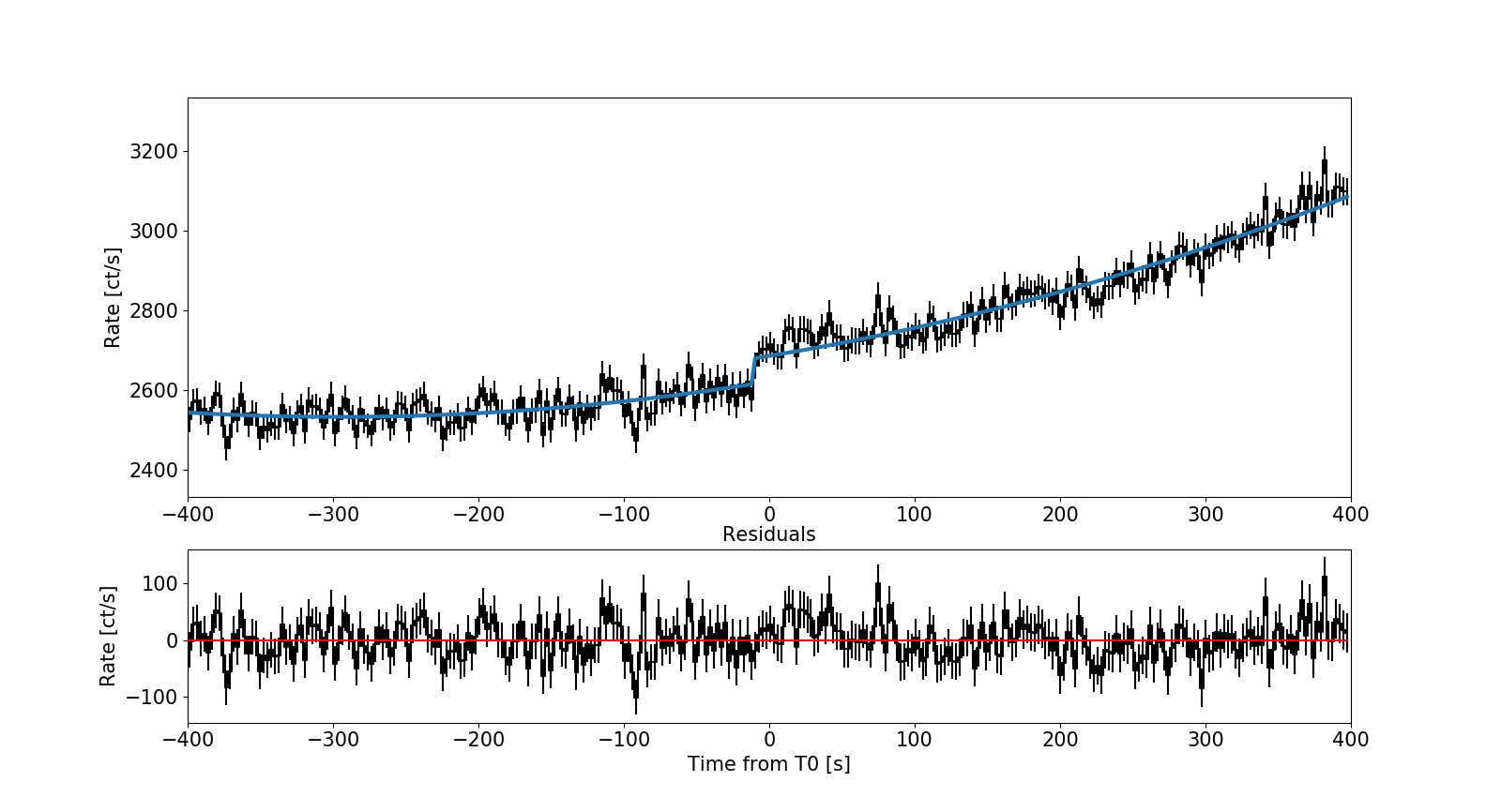}
    \caption{{\em Top panel}: 8-1000 keV lightcurve for FRB\,180725.J0613+67 along with the modelled background (solid line). The step-like increase around the time of the FRB is due to an occultation step of Crab and was modelled using the adjacent orbits. {\em Bottom panel}: residuals of the background modelling.}
\label{fig:fit_aligned}
\end{figure*}

To properly model the step in a data-independent way, we followed this strategy:
\begin{enumerate}
    \item we extracted the lightcurves of the orbits preceding and following the one of the FRB;
    \item we aligned the two lightcurves making the two Crab occultation steps coincide temporally;
    \item we averaged the two lightcurves;
    \item we used a polynomial+step to fit the mean lightcurve using the step amplitude as a free parameter;
    \item we fitted the FRB180725.J0613+67 lightcurve with the same function forcing the step amplitude to be equal to that obtained from the mean of the adjacent orbits.
\end{enumerate}
The result is shown in Figure~\ref{fig:fit_aligned}. This event was included in our final sample as it passed our filtering procedure.

\section{Results}
\label{sec:res}

\begin{figure*}
\centering
\includegraphics[width=\linewidth]{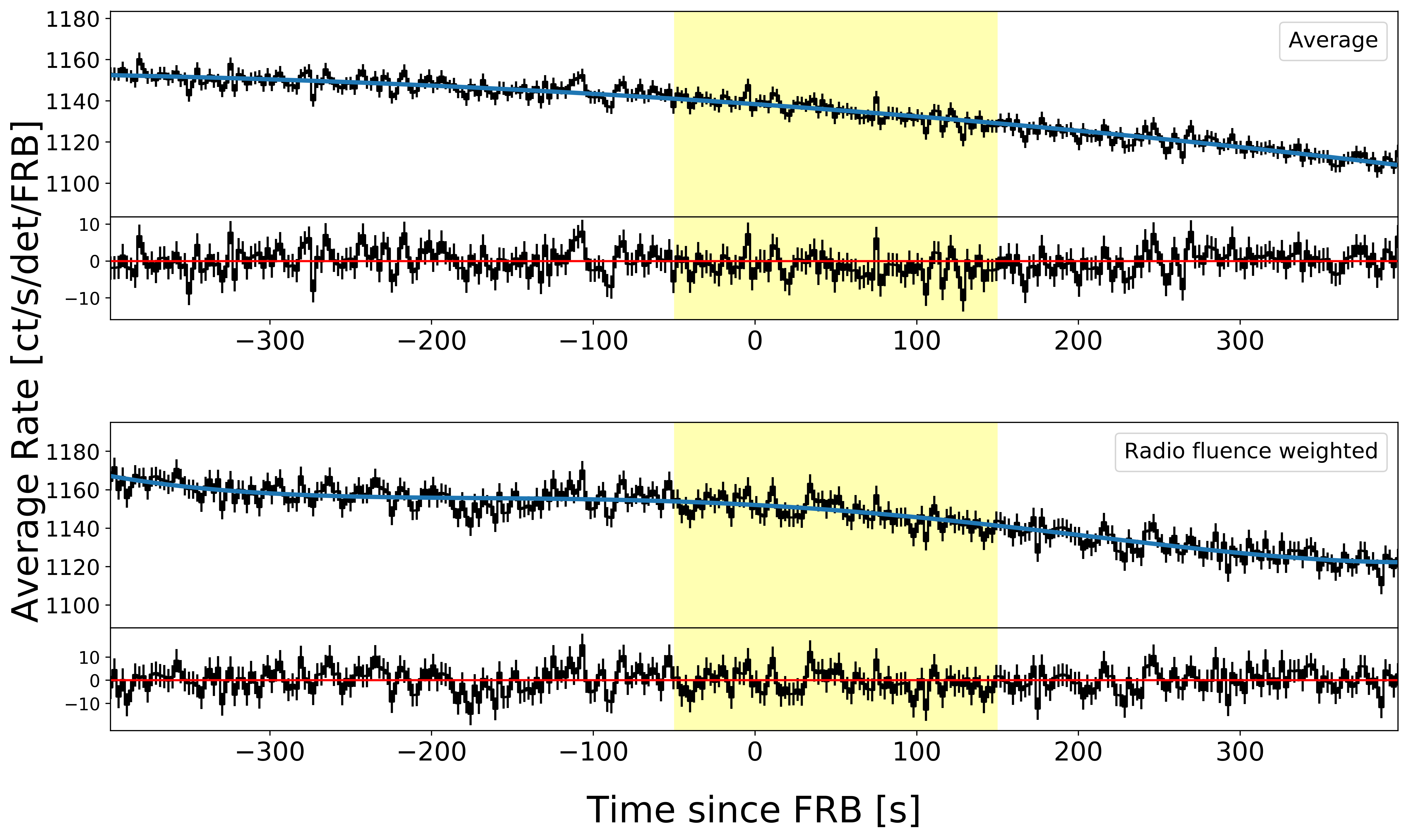}
\caption{Cumulative lightcurves, background fitting and residuals in the 8-1000 keV energy band. In the upper panel we report the result in the case of the arithmetic average of individual lightcurves, while in the lower panel the individual lightcurves are weighted with the radio fluence of the corresponding FRB. The polynomial degrees are respectively 2 and 4.}
\label{fig:combined_soft}
\end{figure*}

\begin{figure*}
\centering
\includegraphics[width=\linewidth]{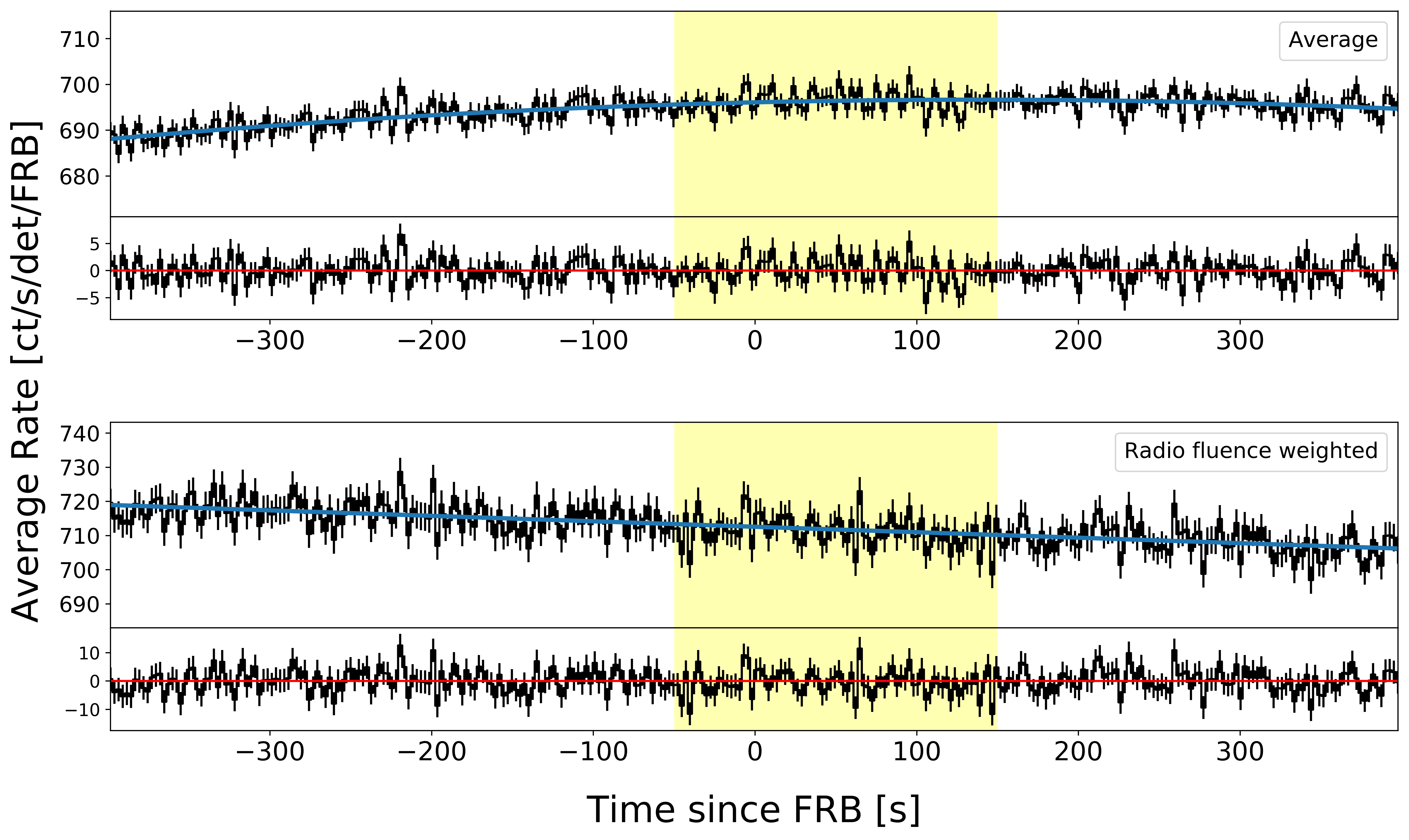}
\caption{Cumulative lightcurves, background fitting and residuals in the 27-1000 keV energy band. In the upper panel we report the result in the case of the arithmetic average of individual lightcurves, while in the lower panel the individual lightcurves are weighted with the radio fluence of the corresponding FRB.  The polynomial degrees are respectively 2 and 1.}
\label{fig:combined_hard}
\end{figure*}

\begin{table}
\centering
\caption{5--$\sigma$ upper-limits on fluence for the power--law and thermal bremsstrahlung spectral shapes.}
\begin{tabular}{ccccc}
\hline
\hline
mode$^\mathrm{a}$ & passband$^\mathrm{b}$ & window & PL$^\mathrm{c}$ & TB$^\mathrm{d}$ \\
\hline
 & & [$\mathrm{s}$] & $^\mathrm{e}$ & $^\mathrm{e}$ \\
\hline
AA & TB & [-50,150] & 4.0 & 1.0 \\
AA & HB & [-50,150] & 3.2 & 1.4 \\
WA & TB & [-50,150] & 6.1 & 1.6 \\
WA & HB & [-50,150] & 6.4 & 2.8 \\
AA & TB & [0,150] & 3.4 & 0.89 \\
AA & HB & [0,150] & 2.9 & 1.2 \\
WA & TB & [0,150] & 5.4 & 1.4 \\
WA & HB & [0,150] & 5.5 & 2.4 \\
AA & TB & [0,100] & 2.8 & 0.74 \\
AA & HB & [0,100] & 2.3 & 1.0 \\
WA & TB & [0,100] & 4.4 & 1.2 \\
WA & HB & [0,100] & 4.5 & 1.9 \\
AA & TB & [0,50] & 2.0 & 0.52 \\
AA & HB & [0,50] & 1.6 & 0.69 \\
WA & TB & [0,50] & 3.1 & 0.82 \\
WA & HB & [0,50] & 3.2 & 1.4 \\
AA & SB & [0,1] & 0.43 & 0.11 \\
AA & HB & [0,1] & 0.36 & 0.15 \\
WA & SB & [0,1] & 0.71 & 0.19 \\
WA & HB & [0,1] & 0.71 & 0.31 \\
\hline
\end{tabular}

\begin{list}{}{}
    \item [$^\mathrm{b}$] AA: Arithmetic Average; WA: Weighted Average.
    \item [$^\mathrm{a}$] TB: Total Band (8-1000 keV); HB: Hard Band (27-1000 keV).
    \item [$^\mathrm{c}$] Power--law spectra, assuming $\Gamma$=1.16.
    \item [$^\mathrm{d}$] Thermal bremsstrahlung spectra, assuming T=75 keV.
    \item [$^\mathrm{e}$] in units of $10^{-7}\mathrm{erg\,cm^{-2}}$.
\end{list}
\label{tab:upper}
\end{table}

\begin{table}
\centering
\caption{5--$\sigma$ upper-limits on fluence for the power--law and thermal bremsstrahlung spectral shapes for the ASKAP bursts.}
\begin{tabular}{ccccc}
\hline
\hline
mode$^\mathrm{a}$ & passband$^\mathrm{b}$ & window & PL$^\mathrm{c}$ & TB$^\mathrm{d}$ \\
\hline
 &  & [$\mathrm{s}$] & $^\mathrm{e}$ & $^\mathrm{e}$ \\
\hline
AA & TB & [-50,150] & 11 & 2.9 \\
AA & HB & [-50,150] & 6.3 & 2.7 \\
WA & TB & [-50,150] & 12 & 3.2 \\
WA & HB & [-50,150] & 8.0 & 3.4 \\
WA & TB & [0,150] & 9.4 & 2.5 \\
WA & HB & [0,150] & 5.5 & 2.4 \\
WA & TB & [0,150] & 11 & 2.8 \\
WA & HB & [0,150] & 7.0 & 3.0 \\
AA & TB & [0,100] & 7.7 & 2.0 \\
AA & HB & [0,100] & 4.5 & 1.9 \\
WA & TB & [0,100] & 8.7 & 2.3 \\
WA & HB & [0,100] & 5.7 & 2.4 \\
AA & TB & [0,50] & 5.5 & 1.5 \\
AA & HB & [0,50] & 3.2 & 1.4 \\
WA & TB & [0,50] & 6.1 & 1.6 \\
WA & HB & [0,50] & 4.1 & 1.8 \\
AA & TB & [0,1] & 1.3 & 0.33 \\
AA & HB & [0,1] & 0.71 & 0.31 \\
WA & TB & [0,1] & 1.4 & 0.37 \\
WA & HB & [0,1] & 0.89 & 0.38 \\
\hline
\end{tabular}

\begin{list}{}{}
    \item [$^\mathrm{b}$] AA: Arithmetic Average; WA: Weighted Average.
    \item [$^\mathrm{a}$] TB: Total Band (8-1000 keV); HB: Hard Band (27-1000 keV).
    \item [$^\mathrm{c}$] Power--law spectra, assuming $\Gamma$=1.16.
    \item [$^\mathrm{d}$] Thermal bremsstrahlung spectra, assuming T=75 keV.
    \item [$^\mathrm{e}$] in units of $10^{-7}\mathrm{erg\,cm^{-2}}$.
\end{list}

\label{tab:upper_ASKAP}
\end{table}

Following the procedure of Sect.~\ref{s:methodology}, the FRB sample shrank to 15 and 22 elements for the total and the hard energy band, respectively. 
The fit of the average lightcurves led to no  $>3\sigma$ excess in the central interval for both energy ranges.
We also produced a second set of mean lightcurves, considering a weighted average, being the weights the radio fluences reported in Table \ref{tab:general}. The rationale behind this choice is the assumption of a common spectral ratio between radio and hard X/$\gamma$--ray frequencies for all FRBs, although we acknowledge it might be a simplistic assumption given the spectral variability in the radio band observed in multiple bursts from the repeating FRB\,121102 \citep{Spitler16,Gajjar18,Hessels18}.
The results of the radio fluence calculation are reported in the $\mathrm{F}_{R}$ column of Table~\ref{tab:general}: we report an empty entry for FRB\,160608 because there was no fluence information from the catalogue. When we calculated the weighted average lightcurves, we assigned to this event a putative fluence corresponding to the average value of the available sample. No significant excess was found in fluence-weighted case either. The four lightcurves so obtained are displayed in Fig.~\ref{fig:combined_soft} (total band) and Fig.~\ref{fig:combined_hard} (hard band).

We calculated the corresponding $5\sigma$ upper limits on the total fluence in the whole energy {\em Fermi}/GBM band-pass 8-1000 keV (and 27-1000 keV sub-band) and in five different time ranges (Table~\ref{tab:upper}). Following D16, we assumed both a power--law (with $\Gamma = 1.16$) and a bremsstrahlung spectrum (with T = 75 keV), while we calculated the radio-to-$\gamma$-ray fluence ratio $\eta=F_{radio}/F_{\gamma}$ assuming the average radio fluence for the bursts reported in \texttt{frbcat}. The most conservative fluence upper limits for the power--law model for 200~s (1~s) integration time imply ${\eta}>10^{8.0}\; (10^{8.9})$ \,Jy\,ms\,erg$^{-1}$\,cm$^2$, considering the average radio fluence density of our sample ($\sim 58$ Jy\,ms). The same calculation for the bremsstrahlung model yields $\eta>10^{8.3}\; (10^{9.2})$ \,Jy\,ms\,erg$^{-1}$\,cm$^2$ for 200~s (1~s) integration time.

\subsection{ASKAP bursts}
The final sample contains 6 events detected by the Australian SKA Pathfinder (ASKAP; \citealt{ASKAP}). These events are worth an additional, specific analysis, given their likely closer average distance compared to that of the remaining sample \citep{Shannon18}, which translates into stronger constraints on the intrinsic energy with respect to the rest of the known FRB population. The upper limits for this subset are reported in Table~\ref{tab:upper_ASKAP}. Only in two cases (170416, 170428) we could study the background in the full band. 

\section{Discussion}
\label{sec:disc}
\begin{figure}
\centering
\includegraphics[width=\linewidth]{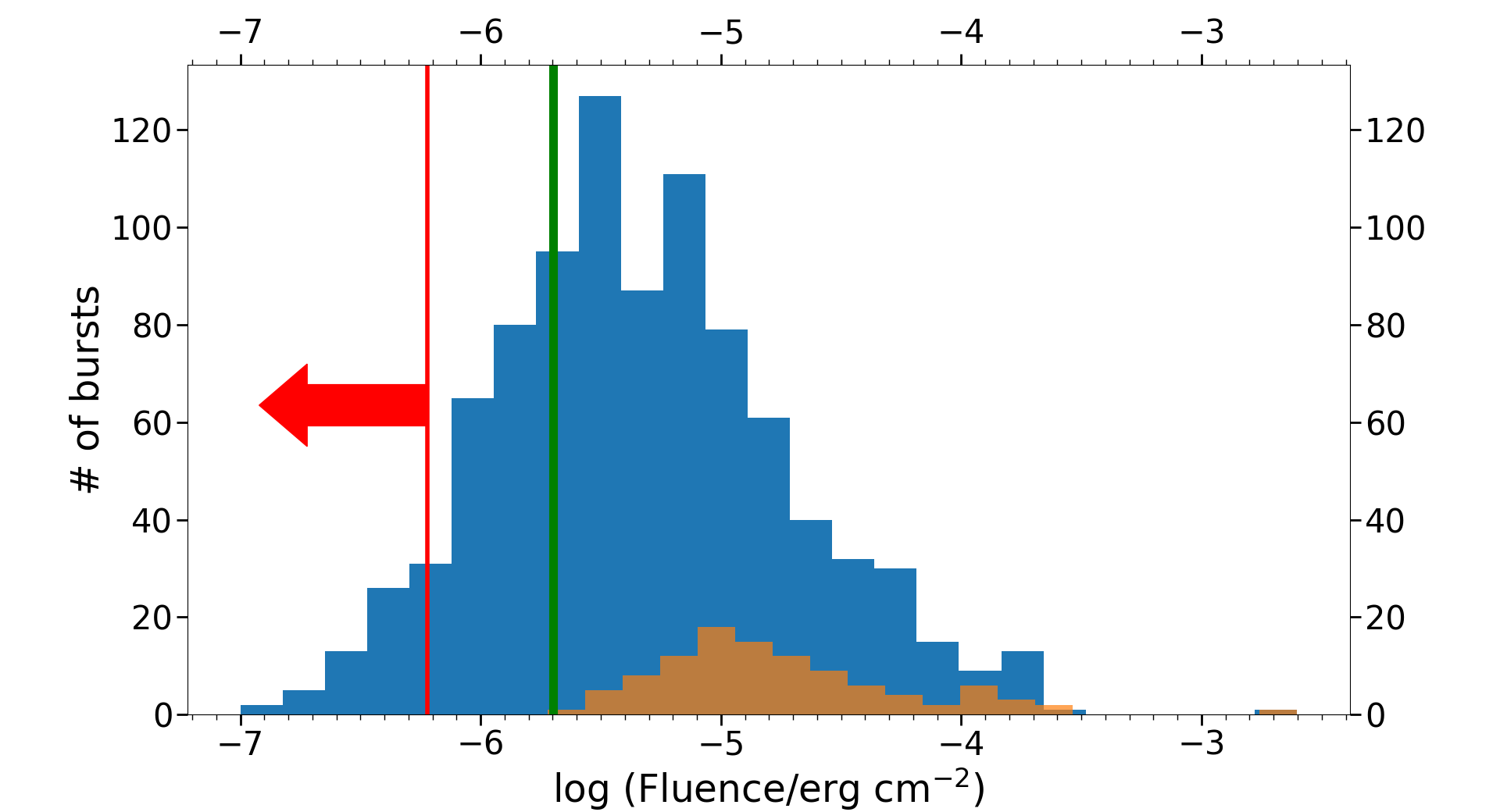}
\caption{200-s fluence limit (red line) in the 8-1000 keV band compared with the distribution of LGRBs ($2\mathrm{s}<T_{90}<100\mathrm{s}$, blue) and very long GRBs ($T_{90}>100\mathrm{s}$, brown) detected by {\em {\em Fermi}/GBM}. The green line represents the quantity $F_{\gamma,D16, resc}$ (see text).}
\label{fig:long}
\end{figure}

\begin{figure}
\centering
\includegraphics[width=\linewidth]{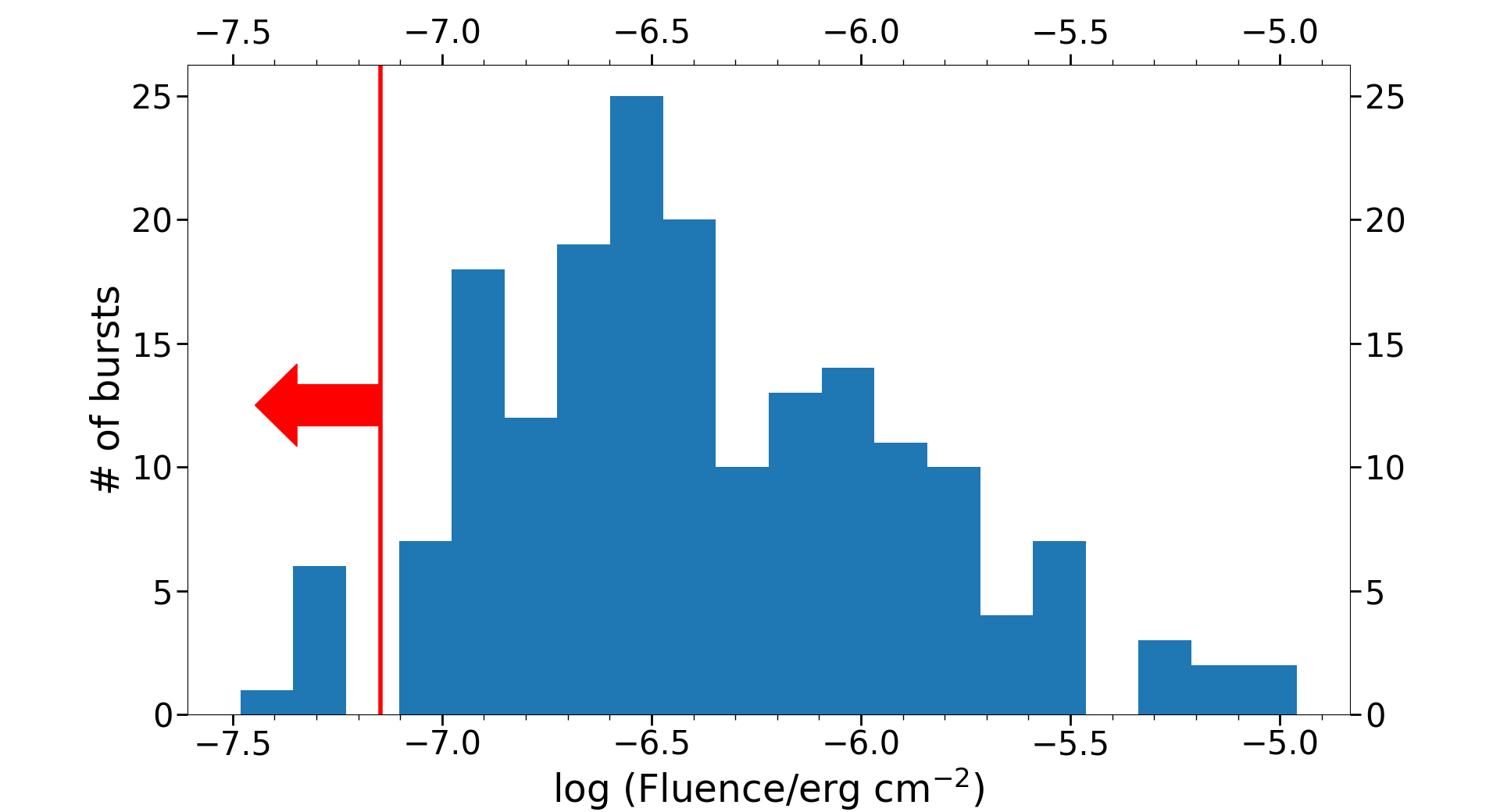}
\caption{1-s fluence limit in the 8-1000 keV band compared with the fluence distribution of SGRBs $T_{90}\leq2\mathrm{s}$ detected by {\em Fermi}/GBM.}
\label{fig:short}
\end{figure}
This work was motivated by the controversial association proposed by D16 between FRB\,131104 and a $\gamma$--ray signal positionally and temporally coincident with the radio burst.
The detection was marginal ($3.2\,\sigma$ confidence) and the analysis revealed a prolonged emission ($T_{90}=377\pm24$~s), with a $\gamma$--ray fluence $F_{\gamma, \mathrm{D16}} = 4 \times 10^{-6}$ erg $\mathrm{cm}^{-2}$ in the $15$--$150$ keV band. 

Our approach was specifically conceived to investigate the presence of a comparably long emission in the {\em Fermi}/GBM data.
Given the highly-variable background of this instrument on long timescales, the entire $\sim400$ s-long window turned out to be too difficult to model, so we analysed a 200 s-long window around the time of the FRB. For a proper comparison, the FRB\,131104 $\gamma$--ray fluence should be rescaled on the 200~s window. Given the uncertain emission history, we assumed a constant luminosity over the entire $\sim400$-s window, finding  $F_{\gamma, \mathrm{D16,resc}} \approx 2 \times 10^{-6}$ erg $\mathrm{cm}^{-2}$ in the 15-150 keV band for the 200-s window. Our upper limits on the same integration time were $\approx$ a few $\times 10^{-7}$~erg\,cm$^{-2}$ in all cases in the 8-1000 keV band. In particular, the most conservative upper limits were obtained assuming a power--law spectrum: the value of $6.4\times10^{-7}$~erg\,cm$^{-2}$ over 200-s integration time is a factor $\sim4$ lower than $F_{\gamma, \mathrm{D16,resc}}$. These limits are even stronger when a thermal bremsstrahlung spectrum is assumed, the most conservative value being $2.8\times10^{-7}$~erg\,cm$^{-2}$, almost one order of magnitude lower than $F_{\gamma, \mathrm{D16,resc}}$. Thus we can exclude that a signal with the same $\gamma$--ray fluence as that found by D16 is a common feature among FRBs. 

Concerning the connection with other astrophysical transients, we tested two kinds of scenarios:
\begin{enumerate}
    \item extragalactic magnetar giant flares;
    \item gamma--ray bursts (GRBs).
\end{enumerate}

Magnetar giant flares are energetic ($3\times10^{44}$-- $2\times10^{46}$~erg) and short-lived (100 ms) events that peak in the hard X--ray band and can be observed up to 30 Mpc (e.g. \citet{Mazets79,Feroci01,Palmer05,Hurley05,Svinkin15}). For galactic magnetars, a dimmer tail (a factor $\sim 10^{-3}$ the peak value) extending up to hundreds of seconds has also been observed.
\citet[hereafter T16]{Tendulkar16} explored the possible connection between these giant-flares and FRBs estimating the radio-to-gamma fluence ratio $\eta$ for both classes of sources. Using the radio fluence upper limits for the Galactic source SGR\,1806-20, they found $\eta_{\rm SGR}\la10^{7}$\,Jy\,ms\,erg$^{-1}$\,cm$^2$ and, using  $\gamma$--ray fluence upper limits for 15 FRBs, $\eta_{FRB}\ga10^{7-9}$\,Jy\,ms\,erg$^{-1}$\,cm$^2$. Giving the incompatibility of the two, T16 concluded that the two kinds of events cannot be associated.
In this regard, we used our 1-s fluence upper limits to further constrain $\eta_{FRB}$. Our most conservative result $\eta_{FRB}>10^{8.9}$ \,Jy\,ms\,erg$^{-1}$\,cm$^2$ (obtained for the power-law spectral shape) is incompatible with $\eta_{\rm SGR}\la10^{7}$\,Jy\,ms\,erg$^{-1}$\,cm$^2$ found for SGR\,1806-20,  suggesting FRBs are not produced by the same mechanism(s) powering giant flares from extragalactic magnetars.

The possible FRB-GRB connection can be explored using our fluence upper limits on both the prolonged (200 s) and short (1 s) timescales.
Figure~\ref{fig:long} shows the fluence distribution of the GBM catalog of LGRBs \citep{Bhat16}, compared with our most conservative limit on the prolonged emission (200-s upper limit): we exclude $\sim94\%$ of the Long GRBs (LGRBs, $T_{90}>2$~s) and {\em all} the very long GRBs ($T_{90}>100$~s).
Finally, we compared our 1-s limits with the fluence distribution of {\em Fermi}/GBM Short GRBs (SGRBs, $T_{90}<2$\,s). Figure~\ref{fig:short} shows that $\sim96\%$ of SGRBs are incompatible with our most conservative upper limit. To conclude, we reject the possibility that the simultaneous presence of a LGRB or a SGRB is a common feature among FRBs, even if a connection in terms of a common progenitor is not excluded \citep{Margalit19}.

\section{Conclusions}
\label{sec:conc}
FRB\,131104 is the first FRB for which a positional and temporal-coincident $\gamma$--ray signal possibly associated has been observed.
This association has been questioned on several grounds and remains controversial. We investigated if the presence of a $\gamma$--ray signal is a common feature among {\em Fermi}/GBM-detected FRBs. Given the wide sky coverage (all the not-Earth-occulted sky, i.e. $\sim 1/3$ of the whole sky) and the extended band-pass (8-1000 keV), {\em Fermi}/GBM continuous data offer the possibility for a stringent test.

Compared to previous analogous works, our approach was innovative for two reasons:
\begin{enumerate}
    \item we investigated a long ($200$ s) time interval around the radio signal. This turned out to be challenging, given the highly-variable {\em Fermi}/GBM background due to its $\sim~26^{\circ}$-inclined orbit. This forced us to develop a background modelling strategy based on an approach like that used in machine learning methodologies. This allowed for the investigation of a previously unexplored extended interval that is crucial to search for faint and comparably long emission.
    \item we carried out a sample study, building and studying cumulative lightcurves, thus improving the sensitivity under the assumption of a common behaviour.
\end{enumerate}

We investigated the $\gamma$--ray lightcurves over four long integration times (200, 150, 100, and 50 s), finding no signal down to $\sim 10^{-7}$ erg cm$^{-1}$ (5--$\sigma$ limit): this excludes an emission like that reported by D16 to be a common feature among FRBs. Furthermore, we analysed the presence of a short-lived (1~s) signal, obtaining a 5--$\sigma$ upper limit of $\sim 10^{-8}$ erg cm $^{-2}$.

Our cumulative approach is complementary to that adopted in other similar works, which searched for high-energy counterparts in individual FRBs.

Concerning the possible FRB-LGRB connection, our most conservative limit excludes the systematic presence of the $94\%$ of {\em Fermi}/GBM-detected LGRBs ($2\mathrm{s}<T_{90}<100$s) and the systematic presence of the entire very long GRBs ($T_{90}>100$s) population.
On the other hand, the possible simultaneous emission from a SGRB is constrained using our upper-limits on the shortest integration time. Our 1-s upper limit excludes the $96\%$ of {\em Fermi}/GBM-detected SGRBs ($T_{90}<2$s) to be a common FRB-associated feature.
Moreover, our limits on the radio-to-gamma fluence ratio $\eta$ ($\eta_{FRB}>10^{8.9}$ \,Jy\,ms\,erg$^{-1}$\,cm$^2$ for 1 s integration time) suggest a different emission process powering FRBs and giant flares from extragalactic magnetars, thus corroborating previous results obtained on smaller samples.

Our analysis benefits from the collective properties of the publicly available FRB catalogue, so it can be further extended to future richer samples and powered by the implementation of our algorithm in a software tool (e.g. training a neural network). Given that the number of FRB discoveries is dramatically rising thanks to the new and forthcoming radio facilities, a (conservative) rate of $\sim1000$ $\mathrm{year}^{-1}$ has been suggested \citep{Petroff19_rev, Lorimer18}. Given that we could use $\sim 1/3$ of the catalogued FRBs, the limits we posed in this paper could be further lowered by a factor $\sim 7$ in a matter of just a few years.
\bibliographystyle{aa}   
\bibliography{alles_grbs}

\begin{thebibliography}{34}
\expandafter\ifx\csname natexlab\endcsname\relax\def\natexlab#1{#1}\fi

\bibitem[{{Atwood} {et~al.}(2009){Atwood}, {Abdo}, {Ackermann}, {Althouse},
  {Anderson}, {Axelsson}, {Baldini}, {Ballet}, {Band}, {Barbiellini},
  {Bartelt}, {Bastieri}, {Baughman}, {Bechtol}, {B{\'e}d{\'e}r{\`e}de},
  {Bellardi}, {Bellazzini}, {Berenji}, {Bignami}, {Bisello}, {Bissaldi},
  {Blandford}, {Bloom}, {Bogart}, {Bonamente}, {Bonnell}, {Borgland },
  {Bouvier}, {Bregeon}, {Brez}, {Brigida}, {Bruel}, {Burnett}, {Busetto},
  {Caliandro}, {Cameron}, {Caraveo}, {Carius}, {Carlson}, {Casandjian},
  {Cavazzuti}, {Ceccanti}, {Cecchi}, {Charles}, {Chekhtman}, {Cheung},
  {Chiang}, {Chipaux}, {Cillis}, {Ciprini}, {Claus}, {Cohen-Tanugi},
  {Condamoor}, {Conrad}, {Corbet}, {Corucci}, {Costamante}, {Cutini}, {Davis},
  {Decotigny}, {DeKlotz}, {Dermer}, {de Angelis}, {Digel}, {do Couto e Silva},
  {Drell}, {Dubois}, {Dumora}, {Edmonds}, {Fabiani}, {Farnier}, {Favuzzi},
  {Flath}, {Fleury}, {Focke}, {Funk}, {Fusco}, {Gargano}, {Gasparrini},
  {Gehrels}, {Gentit}, {Germani}, {Giebels}, {Giglietto}, {Giommi}, {Giordano},
  {Glanzman}, {Godfrey}, {Grenier}, {Grondin}, {Grove}, {Guillemot}, {Guiriec},
  {Haller}, {Harding}, {Hart}, {Hays}, {Healey}, {Hirayama}, {Hjalmarsdotter},
  {Horn}, {Hughes}, {J{\'o}hannesson}, {Johansson}, {Johnson}, {Johnson},
  {Johnson}, {Johnson}, {Kamae}, {Katagiri}, {Kataoka}, {Kavelaars}, {Kawai},
  {Kelly}, {Kerr}, {Klamra}, {Kn{\"o}dlseder}, {Kocian}, {Komin}, {Kuehn},
  {Kuss}, {Landriu}, {Latronico}, {Lee}, {Lee}, {Lemoine-Goumard}, {Lionetto},
  {Longo}, {Loparco}, {Lott}, {Lovellette}, {Lubrano}, {Madejski}, {Makeev},
  {Marangelli}, {Massai}, {Mazziotta}, {McEnery}, {Menon}, {Meurer},
  {Michelson}, {Minuti}, {Mirizzi}, {Mitthumsiri}, {Mizuno}, {Moiseev},
  {Monte}, {Monzani}, {Moretti}, {Morselli}, {Moskalenko}, {Murgia},
  {Nakamori}, {Nishino}, {Nolan}, {Norris}, {Nuss}, {Ohno}, {Ohsugi}, {Omodei},
  {Orlando}, {Ormes}, {Paccagnella}, {Paneque}, {Panetta}, {Parent}, {Pearce},
  {Pepe}, {Perazzo}, {Pesce-Rollins}, {Picozza}, {Pieri}, {Pinchera}, {Piron},
  {Porter}, {Poupard}, {Rain{\`o}}, {Rando}, {Rapposelli}, {Razzano}, {Reimer},
  {Reimer}, {Reposeur}, {Reyes}, {Ritz}, {Rochester}, {Rodriguez}, {Romani},
  {Roth}, {Russell}, {Ryde}, {Sabatini}, {Sadrozinski}, {Sanchez}, {Sand er},
  {Sapozhnikov}, {Parkinson}, {Scargle}, {Schalk}, {Scolieri}, {Sgr{\`o}},
  {Share}, {Shaw}, {Shimokawabe}, {Shrader}, {Sierpowska-Bartosik}, {Siskind},
  {Smith}, {Smith}, {Spandre}, {Spinelli}, {Starck}, {Stephens}, {Strickman},
  {Strong}, {Suson}, {Tajima}, {Takahashi}, {Takahashi}, {Tanaka}, {Tenze},
  {Tether}, {Thayer}, {Thayer}, {Thompson}, {Tibaldo}, {Tibolla}, {Torres},
  {Tosti}, {Tramacere}, {Turri}, {Usher}, {Vilchez}, {Vitale}, {Wang},
  {Watters}, {Winer}, {Wood}, {Ylinen}, \& {Ziegler}}]{Atwood09}
{Atwood}, W.~B., {Abdo}, A.~A., {Ackermann}, M., {et~al.} 2009, \apj, 697, 1071

\bibitem[{{Bannister} {et~al.}(2019){Bannister}, {Deller}, {Phillips},
  {Macquart}, {Prochaska}, {Tejos}, {Ryder}, {Sadler}, {Shannon}, {Simha},
  {Day}, {McQuinn}, {North-Hickey}, {Bhandari}, {Arcus}, {Bennert}, {Burchett},
  {Bouwhuis}, {Dodson}, {Ekers}, {Farah}, {Flynn}, {James}, {Kerr}, {Lenc},
  {Mahony}, {O{\textquoteright}Meara}, {Os{\l}owski}, {Qiu}, {Treu}, {U},
  {Bateman}, {Bock}, {Bolton}, {Brown}, {Bunton}, {Chippendale}, {Cooray},
  {Cornwell}, {Gupta}, {Hayman}, {Kesteven}, {Koribalski}, {MacLeod},
  {McClure-Griffiths}, {Neuhold}, {Norris}, {Pilawa}, {Qiao}, {Reynolds},
  {Roxby}, {Shimwell}, {Voronkov}, \& {Wilson}}]{Bannister19}
{Bannister}, K.~W., {Deller}, A.~T., {Phillips}, C., {et~al.} 2019, Science,
  365, 565

\bibitem[{{Barthelmy} {et~al.}(2005){Barthelmy}, {Barbier}, {Cummings},
  {Fenimore}, {Gehrels}, {Hullinger}, {Krimm}, {Markwardt}, {Palmer},
  {Parsons}, {Sato}, {Suzuki}, {Takahashi}, {Tashiro}, \&
  {Tueller}}]{Barthelmy05}
{Barthelmy}, S.~D., {Barbier}, L.~M., {Cummings}, J.~R., {et~al.} 2005, Space
  Sci. Rev., 120, 143

\bibitem[{{Bhat} {et~al.}(2016){Bhat}, {Meegan}, {von Kienlin}, {Paciesas},
  {Briggs}, {Burgess}, {Burns}, {Chaplin}, {Cleveland}, {Collazzi},
  {Connaughton}, {Diekmann}, {Fitzpatrick}, {Gibby}, {Giles}, {Goldstein},
  {Greiner}, {Jenke}, {Kippen}, {Kouveliotou}, {Mailyan}, {McBreen}, {Pelassa},
  {Preece}, {Roberts}, {Sparke}, {Stanbro}, {Veres}, {Wilson-Hodge}, {Xiong},
  {Younes}, {Yu}, \& {Zhang}}]{Bhat16}
{Bhat}, P., {Meegan}, C.~A., {von Kienlin}, A., {et~al.} 2016, ApJS, 223, 28

\bibitem[{{Chatterjee} {et~al.}(2017){Chatterjee}, {Law}, {Wharton},
  {Burke-Spolaor}, {Hessels}, {Bower}, {Cordes}, {Tendulkar}, {Bassa},
  {Demorest}, {Butler}, {Seymour}, {Scholz}, {Abruzzo}, {Bogdanov}, {Kaspi},
  {Keimpema}, {Lazio}, {Marcote}, {McLaughlin}, {Paragi}, {Ransom}, {Rupen},
  {Spitler}, \& {van Langevelde}}]{Chatterjee17}
{Chatterjee}, S., {Law}, C.~J., {Wharton}, R.~S., {et~al.} 2017, \nat, 541, 58

\bibitem[{{Cunningham} {et~al.}(2019){Cunningham}, {Cenko}, {Burns},
  {Goldstein}, {Lien}, {Kocevski}, {Briggs}, {Connaughton}, {Miller},
  {Racusin}, \& {Stanbro}}]{Cunningham19}
{Cunningham}, V., {Cenko}, S.~B., {Burns}, E., {et~al.} 2019, \apj, 879, 40

\bibitem[{{DeLaunay} {et~al.}(2016){DeLaunay}, {Fox}, {Murase},
  {M{\'e}sz{\'a}ros}, {Keivani}, {Messick}, {Mostaf{\'a}}, {Oikonomou}, {Te{\v
  s}i{\'c}}, \& {Turley}}]{DeLaunay16}
{DeLaunay}, J.~J., {Fox}, D.~B., {Murase}, K., {et~al.} 2016, \apjl, 832, L1

\bibitem[{{Feroci} {et~al.}(2001){Feroci}, {Hurley}, {Duncan}, \&
  {Thompson}}]{Feroci01}
{Feroci}, M., {Hurley}, K., {Duncan}, R.~C., \& {Thompson}, C. 2001, \apj, 549,
  1021

\bibitem[{{Gajjar} {et~al.}(2018){Gajjar}, {Siemion}, {Price}, {Law},
  {Michilli}, {Hessels}, {Chatterjee}, {Archibald}, {Bower}, {Brinkman},
  {Burke-Spolaor}, {Cordes}, {Croft}, {Enriquez}, {Foster}, {Gizani},
  {Hellbourg}, {Isaacson}, {Kaspi}, {Lazio}, {Lebofsky}, {Lynch}, {MacMahon},
  {McLaughlin}, {Ransom}, {Scholz}, {Seymour}, {Spitler}, {Tendulkar},
  {Werthimer}, \& {Zhang}}]{Gajjar18}
{Gajjar}, V., {Siemion}, A.~P.~V., {Price}, D.~C., {et~al.} 2018, \apj, 863, 2

\bibitem[{{Gehrels} {et~al.}(2004){Gehrels}, {Chincarini}, {Giommi}, {Mason},
  {Nousek}, {Wells}, {White}, {Barthelmy}, {Burrows}, {Cominsky}, {Hurley},
  {Marshall}, {M{\'e}sz{\'a}ros}, {Roming}, {Angelini}, {Barbier}, {Belloni},
  {Campana}, {Caraveo}, {Chester}, {Citterio}, {Cline}, {Cropper}, {Cummings},
  {Dean}, {Feigelson}, {Fenimore}, {Frail}, {Fruchter}, {Garmire}, {Gendreau},
  {Ghisellini}, {Greiner}, {Hill}, {Hunsberger}, {Krimm}, {Kulkarni}, {Kumar},
  {Lebrun}, {Lloyd-Ronning}, {Markwardt}, {Mattson}, {Mushotzky}, {Norris},
  {Osborne}, {Paczynski}, {Palmer}, {Park}, {Parsons}, {Paul}, {Rees},
  {Reynolds}, {Rhoads}, {Sasseen}, {Schaefer}, {Short}, {Smale}, {Smith},
  {Stella}, {Tagliaferri}, {Takahashi}, {Tashiro}, {Townsley}, {Tueller},
  {Turner}, {Vietri}, {Voges}, {Ward}, {Willingale}, {Zerbi}, \&
  {Zhang}}]{Gehrels04}
{Gehrels}, N., {Chincarini}, G., {Giommi}, P., {et~al.} 2004, ApJ, 611, 1005

\bibitem[{{Hessels} {et~al.}(2018){Hessels}, {Spitler}, {Seymour}, {Cordes},
  {Michilli}, {Lynch}, {Gourdji}, {Archibald}, {Bassa}, {Bower}, {Chatterjee},
  {Connor}, {Crawford}, {Deneva}, {Gajjar}, {Kaspi}, {Keimpema}, {Law},
  {Marcote}, {McLaughlin}, {Paragi}, {Petroff}, {Ransom}, {Scholz}, {Stappers},
  \& {Tendulkar}}]{Hessels18}
{Hessels}, J.~W.~T., {Spitler}, L.~G., {Seymour}, A.~D., {et~al.} 2018,
  arXiv:1811.10748

\bibitem[{{Hurley} {et~al.}(2005){Hurley}, {Boggs}, {Smith}, {Duncan}, {Lin},
  {Zoglauer}, {Krucker}, {Hurford}, {Hudson}, {Wigger}, {Hajdas}, {Thompson},
  {Mitrofanov}, {Sanin}, {Boynton}, {Fellows}, {von Kienlin}, {Lichti}, {Rau},
  \& {Cline}}]{Hurley05}
{Hurley}, K., {Boggs}, S.~E., {Smith}, D.~M., {et~al.} 2005, \nat, 434, 1098

\bibitem[{{Johnston} {et~al.}(2008){Johnston}, {Taylor}, {Bailes}, {Bartel},
  {Baugh}, {Bietenholz}, {Blake}, {Braun}, {Brown}, {Chatterjee}, {Darling},
  {Deller}, {Dodson}, {Edwards}, {Ekers}, {Ellingsen}, {Feain}, {Gaensler},
  {Haverkorn}, {Hobbs}, {Hopkins}, {Jackson}, {James}, {Joncas}, {Kaspi},
  {Kilborn}, {Koribalski}, {Kothes}, {Landecker}, {Lenc}, {Lovell}, {Macquart},
  {Manchester}, {Matthews}, {McClure-Griffiths}, {Norris}, {Pen}, {Phillips},
  {Power}, {Protheroe}, {Sadler}, {Schmidt}, {Stairs}, {Staveley-Smith},
  {Stil}, {Tingay}, {Tzioumis}, {Walker}, {Wall}, \& {Wolleben}}]{ASKAP}
{Johnston}, S., {Taylor}, R., {Bailes}, M., {et~al.} 2008, Experimental
  Astronomy, 22, 151

\bibitem[{{Katz}(2018)}]{Katz18rev}
{Katz}, J.~I. 2018, Progress in Particle and Nuclear Physics, 103, 1

\bibitem[{{Lorimer}(2018)}]{Lorimer18}
{Lorimer}, D.~R. 2018, Nature Astronomy, 2, 860

\bibitem[{{Lorimer} {et~al.}(2007){Lorimer}, {Bailes}, {McLaughlin},
  {Narkevic}, \& {Crawford}}]{Lorimer07}
{Lorimer}, D.~R., {Bailes}, M., {McLaughlin}, M.~A., {Narkevic}, D.~J., \&
  {Crawford}, F. 2007, Science, 318, 777

\bibitem[{{Margalit} {et~al.}(2019){Margalit}, {Berger}, \&
  {Metzger}}]{Margalit19}
{Margalit}, B., {Berger}, E., \& {Metzger}, B.~D. 2019, arXiv e-prints,
  arXiv:1907.00016

\bibitem[{{Mazets} {et~al.}(1979){Mazets}, {Golentskii}, {Ilinskii}, {Aptekar},
  \& {Guryan}}]{Mazets79}
{Mazets}, E.~P., {Golentskii}, S.~V., {Ilinskii}, V.~N., {Aptekar}, R.~L., \&
  {Guryan}, I.~A. 1979, \nat, 282, 587

\bibitem[{{Meegan} {et~al.}(2009){Meegan}, {Lichti}, {Bhat}, {Bissaldi},
  {Briggs}, {Connaughton}, {Diehl}, {Fishman}, {Greiner}, {Hoover}, {van der
  Horst}, {von Kienlin}, {Kippen}, {Kouveliotou}, {McBreen}, {Paciesas},
  {Preece}, {Steinle}, {Wallace}, {Wilson}, \& {Wilson-Hodge}}]{Meegan09}
{Meegan}, C., {Lichti}, G., {Bhat}, P.~N., {et~al.} 2009, ApJ, 702, 791

\bibitem[{{Palaniswamy} {et~al.}(2018){Palaniswamy}, {Li}, \&
  {Zhang}}]{Palaniswamy18}
{Palaniswamy}, D., {Li}, Y., \& {Zhang}, B. 2018, \apj, 854, L12

\bibitem[{{Palmer} {et~al.}(2005){Palmer}, {Barthelmy}, {Gehrels}, {Kippen},
  {Cayton}, {Kouveliotou}, {Eichler}, {Wijers}, {Woods}, {Granot}, {Lyubarsky},
  {Ramirez-Ruiz}, {Barbier}, {Chester}, {Cummings}, {Fenimore}, {Finger},
  {Gaensler}, {Hullinger}, {Krimm}, {Markwardt}, {Nousek}, {Parsons}, {Patel},
  {Sakamoto}, {Sato}, {Suzuki}, \& {Tueller}}]{Palmer05}
{Palmer}, D.~M., {Barthelmy}, S., {Gehrels}, N., {et~al.} 2005, \nat, 434, 1107

\bibitem[{{Petroff} {et~al.}(2016){Petroff}, {Barr}, {Jameson}, {Keane},
  {Bailes}, {Kramer}, {Morello}, {Tabbara}, \& {van Straten}}]{Petroff16}
{Petroff}, E., {Barr}, E.~D., {Jameson}, A., {et~al.} 2016, PASA, 33, e045

\bibitem[{{Petroff} {et~al.}(2019){Petroff}, {Hessels}, \&
  {Lorimer}}]{Petroff19_rev}
{Petroff}, E., {Hessels}, J. W.~T., \& {Lorimer}, D.~R. 2019, The Astronomy and
  Astrophysics Review, 27, 4

\bibitem[{{Platts} {et~al.}(2018){Platts}, {Weltman}, {Walters}, {Tendulkar},
  {Gordin}, \& {Kandhai}}]{Platts19}
{Platts}, E., {Weltman}, A., {Walters}, A., {et~al.} 2018, arXiv e-prints,
  arXiv:1810.05836

\bibitem[{{Ravi} {et~al.}(2019){Ravi}, {Catha}, {D'Addario}, {Djorgovski},
  {Hallinan}, {Hobbs}, {Kocz}, {Kulkarni}, {Shi}, \& {Vedantham}}]{Ravi19b}
{Ravi}, V., {Catha}, M., {D'Addario}, L., {et~al.} 2019, arXiv e-prints,
  arXiv:1907.01542

\bibitem[{{Scholz} {et~al.}(2017){Scholz}, {Bogdanov}, {Hessels}, {Lynch},
  {Spitler}, {Bassa}, {Bower}, {Burke-Spolaor}, {Butler}, {Chatterjee},
  {Cordes}, {Gourdji}, {Kaspi}, {Law}, {Marcote}, {McLaughlin}, {Michilli},
  {Paragi}, {Ransom}, {Seymour}, {Tendulkar}, \& {Wharton}}]{Scholz17}
{Scholz}, P., {Bogdanov}, S., {Hessels}, J.~W.~T., {et~al.} 2017, \apj, 846, 80

\bibitem[{{Scholz} {et~al.}(2016){Scholz}, {Spitler}, {Hessels}, {Chatterjee},
  {Cordes}, {Kaspi}, {Wharton}, {Bassa}, {Bogdanov}, {Camilo}, {Crawford},
  {Deneva}, {van Leeuwen}, {Lynch}, {Madsen}, {McLaughlin}, {Mickaliger},
  {Parent}, {Patel}, {Ransom}, {Seymour}, {Stairs}, {Stappers}, \&
  {Tendulkar}}]{Scholz16}
{Scholz}, P., {Spitler}, L.~G., {Hessels}, J.~W.~T., {et~al.} 2016, \apj, 833,
  177

\bibitem[{{Shannon} {et~al.}(2018){Shannon}, {Macquart}, {Bannister}, {Ekers},
  {James}, {Os{\l}owski}, {Qiu}, {Sammons}, {Hotan}, {Voronkov}, {Beresford},
  {Brothers}, {Brown}, {Bunton}, {Chippendale}, {Haskins}, {Leach},
  {Marquarding}, {McConnell}, {Pilawa}, {Sadler}, {Troup}, {Tuthill},
  {Whiting}, {Allison}, {Anderson}, {Bell}, {Collier}, {G{\"u}rkan}, {Heald},
  \& {Riseley}}]{Shannon18}
{Shannon}, R.~M., {Macquart}, J.~P., {Bannister}, K.~W., {et~al.} 2018, \nat,
  562, 386

\bibitem[{{Shannon} \& {Ravi}(2017)}]{ShannonRavi17}
{Shannon}, R.~M. \& {Ravi}, V. 2017, \apjl, 837, L22

\bibitem[{{Spitler} {et~al.}(2016){Spitler}, {Scholz}, {Hessels}, {Bogdanov},
  {Brazier}, {Camilo}, {Chatterjee}, {Cordes}, {Crawford}, {Deneva}, {Ferdman},
  {Freire}, {Kaspi}, {Lazarus}, {Lynch}, {Madsen}, {McLaughlin}, {Patel},
  {Ransom}, {Seymour}, {Stairs}, {Stappers}, {van Leeuwen}, \&
  {Zhu}}]{Spitler16}
{Spitler}, L.~G., {Scholz}, P., {Hessels}, J.~W.~T., {et~al.} 2016, \nat, 531,
  202

\bibitem[{{Svinkin} {et~al.}(2015){Svinkin}, {Hurley}, {Aptekar},
  {Golenetskii}, \& {Frederiks}}]{Svinkin15}
{Svinkin}, D.~S., {Hurley}, K., {Aptekar}, R.~L., {Golenetskii}, S.~V., \&
  {Frederiks}, D.~D. 2015, \mnras, 447, 1028

\bibitem[{{Tendulkar} {et~al.}(2017){Tendulkar}, {Bassa}, {Cordes}, {Bower},
  {Law}, {Chatterjee}, {Adams}, {Bogdanov}, {Burke-Spolaor}, {Butler},
  {Demorest}, {Hessels}, {Kaspi}, {Lazio}, {Maddox}, {Marcote}, {McLaughlin},
  {Paragi}, {Ransom}, {Scholz}, {Seymour}, {Spitler}, {van Langevelde}, \&
  {Wharton}}]{Tendulkar17}
{Tendulkar}, S.~P., {Bassa}, C.~G., {Cordes}, J.~M., {et~al.} 2017, \apjl, 834,
  L7

\bibitem[{{Tendulkar} {et~al.}(2016){Tendulkar}, {Kaspi}, \&
  {Patel}}]{Tendulkar16}
{Tendulkar}, S.~P., {Kaspi}, V.~M., \& {Patel}, C. 2016, \apj, 827, 59

\bibitem[{{Thornton} {et~al.}(2013){Thornton}, {Stappers}, {Bailes},
  {Barsdell}, {Bates}, {Bhat}, {Burgay}, {Burke-Spolaor}, {Champion}, {Coster},
  {D'Amico}, {Jameson}, {Johnston}, {Keith}, {Kramer}, {Levin}, {Milia}, {Ng},
  {Possenti}, \& {van Straten}}]{Thornton13}
{Thornton}, D., {Stappers}, B., {Bailes}, M., {et~al.} 2013, Science, 341, 53

\end{thebibliography}

\end{document}